\documentclass[svgnames]{aa}
\usepackage{txfonts}

\usepackage{longtable}
\usepackage{lscape}
\usepackage{booktabs}
\usepackage{graphicx}

\newcommand{\Msol}{\,M\ensuremath{_{\sun}}}
\newcommand{\kms}{\ensuremath{\text{km\ s}^{-1}}}

\bibpunct{(}{)}{,}{a}{}{,} 

\begin{document} 

   \title{Convective core sizes in rotating massive stars}

   \subtitle{I. Constraints from solar metallicity OB field stars}
 \author{S. Martinet
          \inst{1}
          \and G. Meynet\inst{1}
          \and S. Ekstr\"om\inst{1}
          \and S. Sim\'on-D\'iaz\inst{2,3}
          \and G.Holgado\inst{4}
          \and N. Castro\inst{5}
          \and C. Georgy\inst{1}
          \and P. Eggenberger\inst{1}
          \and G. Buldgen\inst{1}
          \and S. Salmon\inst{1}          
          \and R. Hirschi\inst{6,7}
          \and J. Groh\inst{8}
          \and E. Farrell\inst{8}
          \and L. Murphy\inst{8}
}

   \institute{Department of astronomy, University of Geneva, chemin Pegasi 51, 1290 Versoix, Switzerland\\
              \email{sebastien.martinet@unige.ch}
        \and Instituto de Astrof\'isica de Canarias, Avenida V\'ia L\'actea, E-38205 La Laguna, Tenerife, Spain
       \and Universidad de La Laguna, Dpto. Astrof\'isica, E-38206 La Laguna, Tenerife, Spain
        \and Centro de Astrobiolog\'ia, ESAC campus, Villanueva de la Ca\~nada, E-28\,692, Spain
        \and Leibniz-Institut für Astrophysik Potsdam, An der Sternwarte 16, 14482 Potsdam, Germany
         \and Astrophysics group, Keele University, Keele, Staffordshire ST5 5BG, UK
                  \and Institute for the Physics and Mathematics of the Universe (WPI), University of Tokyo, 5-1-5 Kashiwanoha, Kashiwa 277-8583, Japan
         \and Trinity College Dublin, The University of Dublin, Dublin, Ireland
             }



 
  \abstract
   {Spectroscopic studies of Galactic O and B stars show that many stars with masses above 8\ \Msol\ have been observed in the Hertzsprung-Russell (HR) diagram just beyond the main-sequence (MS) band, as predicted by stellar models computed with a moderate overshooting. This may be an indication that the convective core sizes in stars in the upper part of the HR diagram are larger than predicted by these models.}
   {
   Combining stellar evolution models and spectroscopic parameters derived for a large sample of Galactic O and B stars with the inclusion of brand-new information about their projected rotational velocities, %
   we reexamine the question of the convective core size in MS massive stars.}
   {We computed a grid of 120 different stellar evolutionary tracks with {three} initial rotations at solar metallicity (Z = 0.014), spanning a mass range from 7 to 25 \Msol, and combining different values for the initial rotation rate and overshooting parameter. For the rotating models, we considered two cases, one with a moderate and one with a strong angular momentum transport, the latter imposing a solid body rotation during most of the MS phase. We confront the results with {two}
   observed features: the position of the terminal age main sequence (TAMS) in the HR diagram and the decrease of the surface rotation when the surface gravity decreases at the end of the MS phase.
} 
   {{We confirm that for stars more massive than about 8 M$_\odot$, the convective core size at the end of the MS phase increases more rapidly with the mass than in models computed with a constant step overshoot chosen to reproduce the main sequence width in the low mass range (around 2 M$_\odot$).}
   This conclusion is valid for both the cases of non-rotating models and rotating models either with a moderate or a strong angular momentum transport. 
   {The increase of the convective core mass with the mass obtained from the TAMS position
   is, however, larger than the one deduced from the surface velocity drop for masses above about 15\Msol. Although the observations that are available at present cannot determine the best choice between the core sizes given by the TAMS and the velocity drop, we discuss various methods of escaping this dilemma. At the moment, comparisons with 
   eclipsing binaries seem to favor the solution given by the velocity drop.} 

 }
   {
   While we confirm the need for larger convective cores at higher masses, we find tensions among different methods {for stars more massive than 15 M$_\odot$}. The use of single-aged stellar populations {(non-interacting binaries or stellar clusters) would be a great asset in resolving this tension}.
   }
\maketitle



\section{Introduction}
The size of convective cores is an long-standing problem that was investigated very early on by, among others, \cite{SaslawSchwarzschild1965}, \cite{ShavivSalpeter1973}, \cite{Maeder1975}, \cite{Roxburgh1978}, and \cite{Bressan1981},  as well as further on by \cite{Renzini1987}, \cite{Maeder&Meynet1987}, and \cite{Zahn1991}. 
The convective boundaries are defined in stellar models by the Schwarzschild or Ledoux criterion. Both criteria give the position inside the star where the acceleration resulting from the net balance between the gravity and the buoyancy force becomes zero \citep{MaederBook2009}.
However, it neglects the inertia of the convective elements, which will pursue their movements beyond the limit where the acceleration is zero up to the point where the velocity is zero.
While the physics of this process is very basic, there are many different ways to implement it in stellar models \citep[see][]{Gabriel2014}. 
{The two most frequently used methods at the moment are} known as ``step overshoot'' as applied, for example, in \cite{StellarGRIDEkstrom2012} {when defining an overshooting parameter, $\alpha_\text{ov}$, extending the convective radius instantaneously by a fraction of the pressure scale height; whereas}
``diffusive overshoot', for example, as implemented in \cite{Herwig1997, Paxton2011, Choi2016}, proposes an overshooting parameter, f$_\text{ov}$, intervening in a depth-dependent diffusion coefficient.

The overshooting process is not the only physical process impacting the size of the convective cores. Effects such as rotation, tides, or even the presence of an internal magnetic field can all affect the sizes of the convective cores.
This complicates the issue and makes it difficult to find a solution.  

In recent decades, various attempts have been made to constrain the size of the convective core using well-observed features: the main-sequence (MS) band width in the Hertzsprung-Russell (HR) diagram \citep{MaederMermilliod1981,Barbaro1984,Castro2014}, the drop of the surface rotation velocity when plotted as a function of the surface gravity \citep{Brott2011}, or even the extension of the blue loops during the core He-burning phase \citep{Matraka1982,Miller2020}. 

\cite{Castro2014} presented, for the first time, the distribution of a large sample of Galactic massive stars in the spectroscopic 
HR diagram \citep[based on the gravity-effective temperature diagram, see][]{LangerKudritzki2014}, and showed that stellar models from \cite{StellarGRIDEkstrom2012} exhibit a good agreement with the MS width for stars around 8\Msol\ 
, while a better agreement for stars around 15\Msol\ 
was found for the models from \cite{Brott2011} with larger convective core sizes.
This led to suggest a possible scaling of the overshooting with mass. \citet{Claret2016, Claret2017, Claret2018, Claret2019} also provided empirical evidence from the study of a sample of eclipsing binaries that there is a clear and steep increase of the overshooting from 1.2 to 2\Msol, and that it remains constant thereafter out to at least 4.4\Msol. {Recently \citet{Higgins2019, Tkachenko2020} used eclipsing binaries 
to constrain stellar models. Using non-rotating stellar models, \citet{Tkachenko2020} find convective core masses between 17 and 35\% of the stellar mass for stars with masses between 5 and 16 M$_\odot$\footnote{In Sect.~5, we compare these convective core mass fractions with those obtained in the present work.}}.

A drop of the surface velocity is expected to occur just after the MS phase when the star rapidly crosses the HR gap and becomes first a blue and then a red supergiant.
By studying a sample of about 400 O- and B-type stars in the Magellanic Clouds, \cite{Hunter2008} observed a steep drop in the distribution of projected rotational velocities as a function of surface gravity ($g$) at $\log(g)\sim$3.2 
and suggested the slower B supergiants could be post-MS stars. \citet{Brott2011} used this steep drop to constrain the core overshooting parameter in massive star models, while \cite{Vink2010} proposed another interpretation of this drop as being due to an increase of the mass loss connected with the effect of 
the so-called bi-stability jump at T$_\text{eff}$ = 22000 K.

Additional constraints are available from asteroseismology studies \citep{Appourchaux2015}. 
For low-mass stars (below 1.6\Msol), \cite{Deheuvels2016} find hints of overshooting scaling with mass. Using the stellar evolution code CESAM2K, implementing a step overshoot, they obtain values for the extension of the convective core size that are similar to the values obtained by \citet{StellarGRIDEkstrom2012} for the non-rotating models. 
\cite{Bossini2017} revealed the need for a moderate overshooting while studying red clump stars in field and open clusters. Inversion techniques \citep{Roxburgh2002,Buldgen2018} are also used to obtain additional constraints.
Slowly pulsating main-sequence B stars between 2.5 and 8 M$_\odot$, $\delta$ Scuti\ /\ $\gamma$ Dor hybrid stars have also been studied to derive the extension of the core \citep{Moravveji2015,Moravveji2016,Szewczuk2018,Murphy2016}.
More massive stars, $\beta$ Cephei stars (typically between 6-8 and 20\Msol) have been studied  by, among others,  \citet{Aerts2003, Dupret&Aerts2004, Handler2005,Mazumdar&Aerts2006,Briquet&Aerts2007}, and have been reviewed by \cite{Aerts2015}.


The range of convective core sizes deduced from these different studies is still large, which may be understood due to the fact that, as mentioned above, many factors, which may differ from star to star, have an impact as the initial mass, the age, the chemical composition, the rotation, {and even on the mass transfer history \citep[see][]{Klencki2020}.}
Moreover, asteroseismological features, as well as the observation of the MS band,  
cannot specifically constrain the physical process determining the convective core size.
To do so, we 
should identify a feature that only depends on this specific process.

Another approach is to use sophisticated 3D hydrodynamical simulations.
These  simulations can guide the building of new prescriptions for determining the sizes of the convective regions in 1D stellar evolution codes\footnote{Three-dimensional stellar models are too computationally expensive for allowing to follow the evolution of the limit of the convective core during even a significant fraction of the MS phase.}. For instance, simulations such as those of \citet{Meakin2007} and \citet{Cristini2019} 
have shown that the upper boundary of a convective zone may evolve in a way that is not predicted by present 1D stellar models through a process called the mixing boundary entrainment. Some first tests are underway, implementing these new prescriptions in 1D stellar models \citep[][]{Scott2021}.




In the present work, we follow a more classical approach consisting of calibrating the size of the convective cores using observed features dependent on it, as done in previous works \citep[see e.g.,][]{Brott2011,StellarGRIDEkstrom2012}. {However, at the moment, very few authors (if any) have considered the use of more than one observed feature when performing such calibrations. This actually may provide a biased view in the sense that confronting the models with only one observational constraint does not guarantee that the others will be fitted too. The present work is an attempt to perform comparisons with two observed features \footnote{Other observed features such as the extension of the blue loops and the chemical enrichment of the stellar surface during the evolution of massive stars could also be incorporated to such investigation (see Sect. 6).}}: (1) the position of the terminal age main sequence (TAMS) line in the HR diagram; (2) the position of the drop of the surface velocity in the $\upsilon\sin i$ versus surface gravity diagram.

    For the above comparisons, we use two types of rotating models. One set of models uses a given physic input exactly similar to the one used in our grids of models \citep{StellarGRIDEkstrom2012}, where during the MS phase, some differential rotation between the core and the envelope develops (moderate angular momentum transport). A second set has been obtained assuming that stars rotate as solid bodies during most of the MS phase (strong angular momentum transport). {This allows us to investigate, within the framework of a given stellar evolution code, the dependence of the size of the convective core on the angular and chemical mixing processes.}

The paper is organized as follows. The physical ingredients of the models are discussed in Sect. \ref{section:Physical Ingredients}. The properties of the models and the impact of overshooting are discussed in Sect. \ref{section:Features stellar models}, while comparisons with massive stars observations is carried out in Sect. \ref{section:Comparaison}. 
We discuss some limitations of the present approach in Sect. \ref{section:Discussion}. We describe our main findings in Sect. \ref{section:Conclusion}.





\section{Physical ingredients and models computed}


\label{section:Physical Ingredients}
A detailed description of the physics included in the models can be found in \cite{StellarGRIDEkstrom2012}. 
Here, we briefly describe two aspects that are relevant for the present work: overshooting and rotation.

\subsection{Overshooting}

In the stellar grids published thus far \citep{StellarGRIDEkstrom2012, grids1p7_15_2013,georgy2014}, we used models computed with the Schwarzschild criterion and a step overshooting.
In these models, the radius of the convective core is obtained in the following way: 
\begin{align}
  R_\text{conv} = R_\text{conv-Sch}+\alpha \times H_p,
\label{equation:step overshooting} 
\end{align}
where $R_\text{conv-Sch}$ is the radius given by the Schwarzschild criterion, $H_p$ is the pressure scale height estimated at the Schwarzschild radius, and $\alpha$ is a free parameter. In all these models, a value of $\alpha=0.1$ was chosen in order to fit the empirical width of the MS obtained for solar metallicity stars with initial masses in the 1.7\,--\,2.5\Msol\ range. 
This value is kept constant for all the initial masses above 1.7\Msol, for the different metallicities and for core H- and He-burning phases.

The implementation, shown in Eq. \ref{equation:step overshooting}, is very simple. However,  here we should recall three points. First, in the whole convective core (i.e., the Schwarzschild core plus the overshooting region), the temperature gradient is the adiabatic one. Second, the chemical species are completely homogenized in the whole core at every time step during the core H and He-burning phases.\footnote{In the more advanced phases of the evolution, when the nuclear burning time of some elements in some deep layers of the convective core becomes shorter than the timescale for the mixing in that core, the complete homogenization is no longer realized. A diffusive approach has then to be used for mixing the elements in the convective regions.} Finally, we do not consider any overshooting at the borders of the intermediate convective shells or at the bottom of the convective envelope. A brief discussion of these limitations is provided in Sect. \ref{section:Discussion}.

It is also important to remember that while penetrative and diffusive overshooting are quite different, the physical properties at the border of the core are quite similar among non-rotating diffusive models and our penetrative rotating models, even for slow rotators \citep[see][]{Miglio2009,salmon2014}. A more in-depth discussion can be found in Sect. \ref{section:Discussion}.

\subsection{Rotation}

The way the rotation physic is included in the Geneva stellar evolution code (GENEC) 
is presented in \citet[][also  see references therein]{Eggenberger2008}. In the case of models with a moderate angular momentum transport, we use the same prescriptions as in \cite{StellarGRIDEkstrom2012}. In particular, we use the shear diffusion coefficient as given by \cite{Maeder1997} and the horizontal diffusion coefficient from \cite{Zahn1992}. An advective equation is solved for describing the transport of the angular momentum by the meridional currents during the MS phase.

We also considered the case of a very efficient angular momentum transport to take into account the need for a more efficient transport process in asteroseismologic studies of subgiants and giants \citep{Deheuvels2012,Deheuvels2015,Mosser2012,denHartogh2019}.
For this, we used a diffusive approach for the angular momentum transport by the meridional currents, with a very large diffusive coefficient as given by \citet{Song2}.
Due to the nearly constant $\Omega$ profile, transport of chemical elements by shear mixing is negligible. On the other hand, their transport by meridional currents is very strong. These models, for a given initial mass, rotation, and at a given age, are much more mixed than the models with a moderate angular momentum transport. 

\subsection{Models computed}

 The grid is composed of 120 different stellar evolutionary tracks at $Z = 0.014$, with initial masses equal to 7, 9, 15, and 25\Msol, with six different overshooting parameters ranging from 0.1 to 0.6, with surface rotation at the equator ($\upsilon$) on the ZAMS equal to 0, 0.2, and 0.4 the critical velocity\footnote{The critical velocity is the value of the equatorial velocity such that the centrifugal force balances the gravity. The critical velocity in the frame of the Roche approximation is given by $\sqrt{\frac{2}{3}\frac{GM}{R_\text{p,crit}}}$, where $R_\text{p,crit}$ is the polar radius at the critical limit.} 
 ($\upsilon_\text{crit}$). Each rotating model 
 is computed with a moderate and a strong angular momentum transport. All the models were computed until at least the maximum redward extension of the track in the HRD during the core H-burning phase ({\it } i.e. TAMS). Actually, except in a few cases (exactly 17 models), all models were computed until the end of the core H-burning phases. Those that were stopped before this point are chemically mixed to a significant extent and do not contribute to the extension of the MS band width beyond its extension obtained from slower or non-rotating models.
 A little more than one-third of the models (42 exactly) were also computed over the whole core He-burning phase. 




\section{Impact of overshooting and rotation}
\label{section:Features stellar models}

\subsection{Hertzsprung-Russell diagram}

\begin{figure*}[!ht]
    \centering
    \includegraphics[width=1\textwidth]{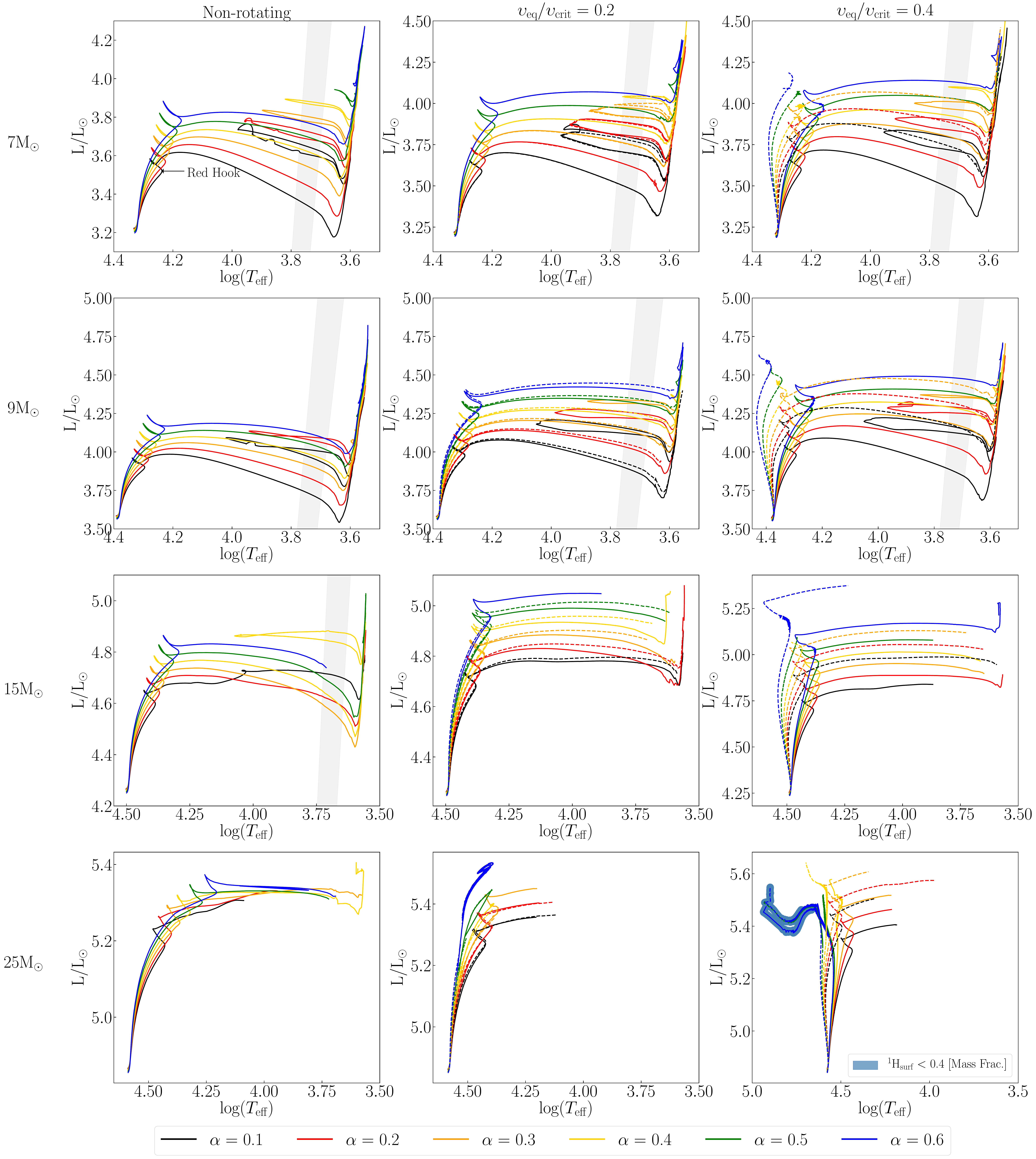}
    \caption{Evolutionary tracks in the Hertzsprung-Russell diagrams. Non-rotating models are shown in the panels in the left column. The plain lines in the panels of the middle and right column are the rotating models with a moderate angular momentum transport. The dashed lines are the models computed with a strong angular momentum transport. We note that at low rotations ($\upsilon/\upsilon_\text{crit}=0.2$), models with moderate and strong angular momentum transport may overlap nearly exactly preventing to see the dashed lines (see e.g., the 7\Msol\ stellar model). 
    The shaded area shows the instability strip \citep[determined as in][]{Anderson2016}. The models were computed up to MS turnoff for most stars, the lowest mass stars were expended up to the end of He-burning. In the bottom right panel, the parts of the tracks where the mass fraction of hydrogen at the surface is below 0.4, meaning the star may be considered a Wolf-Rayet, are highlighted by a broad blue band.
    }
    \label{fig:Grid HRD}
\end{figure*}


Figure~\ref{fig:Grid HRD} presents the evolutionary tracks for the different initial mass models considered in this work, non-rotating and rotating, with various overshoots. 
Let us call the "red hook" the point on the MS band where the tracks reach their lowest effective temperature before turning back, for a short time, to bluer regions of the HR diagram. As the overshooting parameter increases, the red hook occurs in general at a higher luminosity and a lower effective temperature. There are exceptions to this general trend occurring for the most chemically mixed models (see, e.g., the models computed with $\upsilon/\upsilon_\text{crit}=0.4$ and a strong angular momentum transport shown by the dashed lines in the right panels of Fig.~\ref{fig:Grid HRD}).


The blue tracks for the 25\Msol\ in the lower right panel ($\upsilon/\upsilon_\text{crit}=0.4$, both with a moderate and a strong angular momentum transport and an overshooting parameter equal to 0.6), actually show surface abundances during the MS phase that are similar to the surface abundances observed at the surface of Wolf-Rayet stars. Due to our prescriptions for computing the mass-loss rates, much stronger stellar winds appear from that point on and this produces the peculiar shape of the tracks in the regions indicated by the blue shade in the bottom right panel of Fig.~\ref{fig:Grid HRD}.

We can also see the differences in the evolutionary tracks when a moderate and a strong angular momentum transport is considered for a given initial mass, rotation, and overshoot. For the velocities $\upsilon/\upsilon_\text{crit}=0.2$, differences are very small. They are much more important for the cases $\upsilon/\upsilon_\text{crit}=0.4$. A strong coupling implies much more chemically mixed models. This shifts the tracks to the blue with respect to the case of moderate angular momentum transport. 

\begin{figure*}
    \centering
    \includegraphics[width=0.97\textwidth]{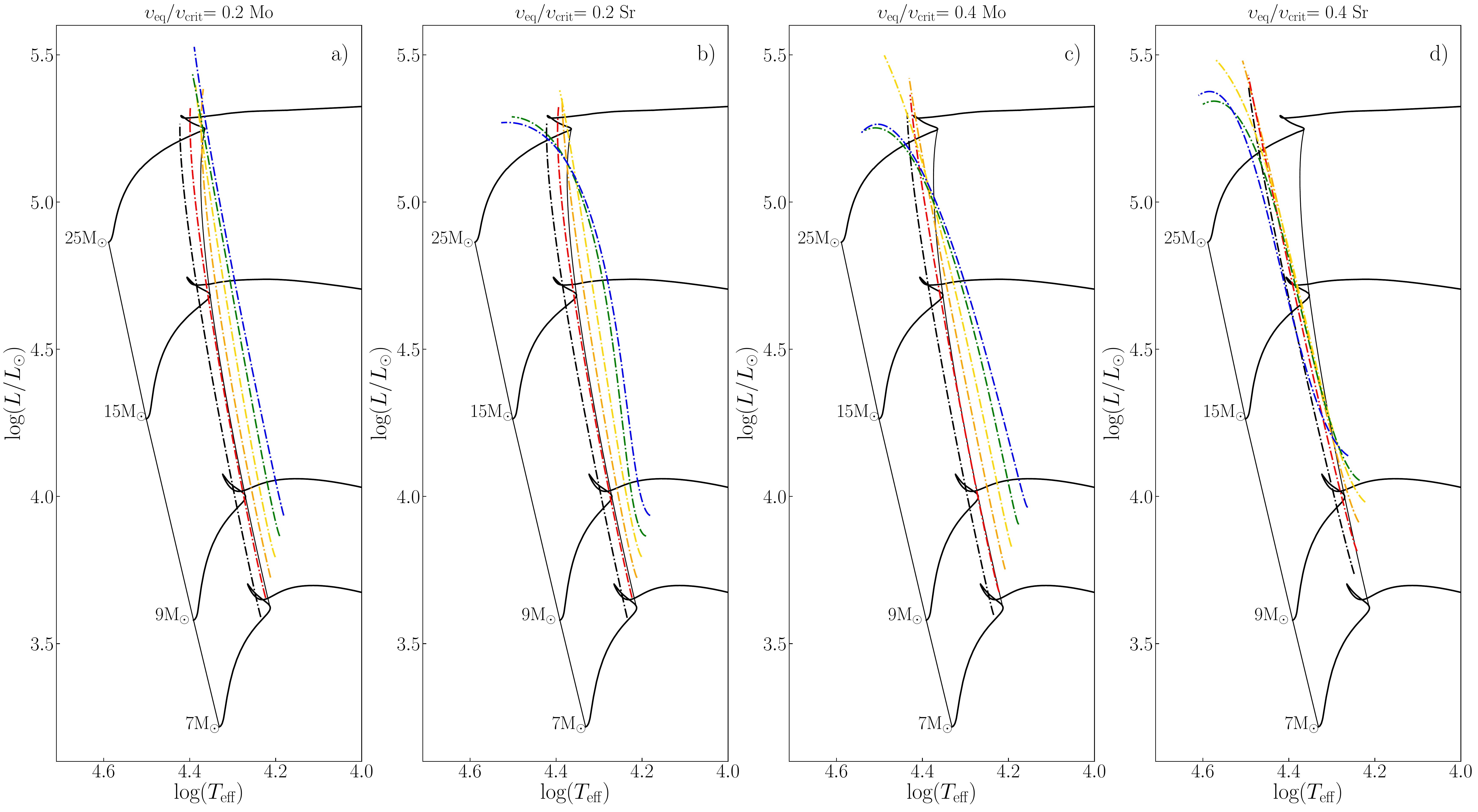}
    \caption{Evolutionary tracks in the HR diagram for non-rotating models with $\alpha$=0.3 (black {solid} lines). The limits of the MS band are indicated for various values of $\alpha$, rotation rates (indicated above each panel), and transport efficiency (see above each panel; Mo/Sr: moderate/strong angular momentum transport). Looking at the bottom of these lines in the left panel, $\alpha$ goes from 0.1 to 0.2, 0.3, 0.4, 0.5, and 0.6, going from left to right.}
    \label{fig:HRD_position_L_max}
\end{figure*}

Figure~\ref{fig:HRD_position_L_max} shows how the width of the MS band varies when different values of $\alpha$ are used in different rotating models. 
We define the MS band width as the difference,
taken at a fixed effective temperature between the luminosity at the red hook and at the ZAMS.
Lines connecting the red hooks of models obtained with the same overshooting parameter can be seen in Fig.~\ref{fig:HRD_position_L_max} (dashed lines except for the case $\alpha=0.3$ which is shown by a {solid} line). This width, in general, increases when the overshooting increases. 

For non-rotating models, at higher masses, the position of the red hook in the HRD extends further away to the red than at lower masses. 
In higher mass models, {due to mass losses by stellar winds}, the core occupies a larger fraction of the total mass at the end of the MS phase and this shifts the TAMS to the red \citep[see e.g.,][]{MaederMeynet1987}.  

On the other hand, in the non-rotating models, enlarging the overshoot continuously increases the MS band width (at least in the range of overshoots considered here); for the rotating models, the impacts between 15 and 25\Msol are very different. A most striking effect can be seen for the 25\Msol\ stellar models at $\upsilon/\upsilon_\text{crit}=0.4$ with moderate or strong angular momentum. In those cases, above some overshoot, any further increase of $\alpha$ produces a reduction of the width of the MS band. This can be understood by the fact that rotation has two counteracting effects on the width of the MS band. On one hand, it slows down the decrease in mass of the convective core during the MS phase. 
On the other hand, rotation also changes the chemical composition in the radiative envelope. 
Rotation, for instance, makes the radiative envelope more helium-rich. This reduces its opacity (massive stars are dominated by electron-scattering opacity) and makes the track bluer, reducing thus the MS band width.
In stars undergoing a strong chemical mixing, this second effect dominates and thus reduces the MS band width compared to analog models without rotation. The more massive a star, the more mixed it is (given an initial rotation, whether computed with a moderate or strong angular momentum transport), thus, this effect appears here only for a mass above a limit between 15 and 25\Msol. It is interesting to compare panels c and d of Fig.~\ref{fig:HRD_position_L_max}. We see that models with a strong angular momentum transport show less sensitivity on the overshoot than models with a moderate angular momentum. This is because models with a strong angular momentum transport are more chemically mixed, making these models less sensitive to larger cores. 

\subsection{Surface rotation}
\label{subsection:surface velocity}

\begin{figure}
    \centering
    \includegraphics[width=0.48\textwidth]{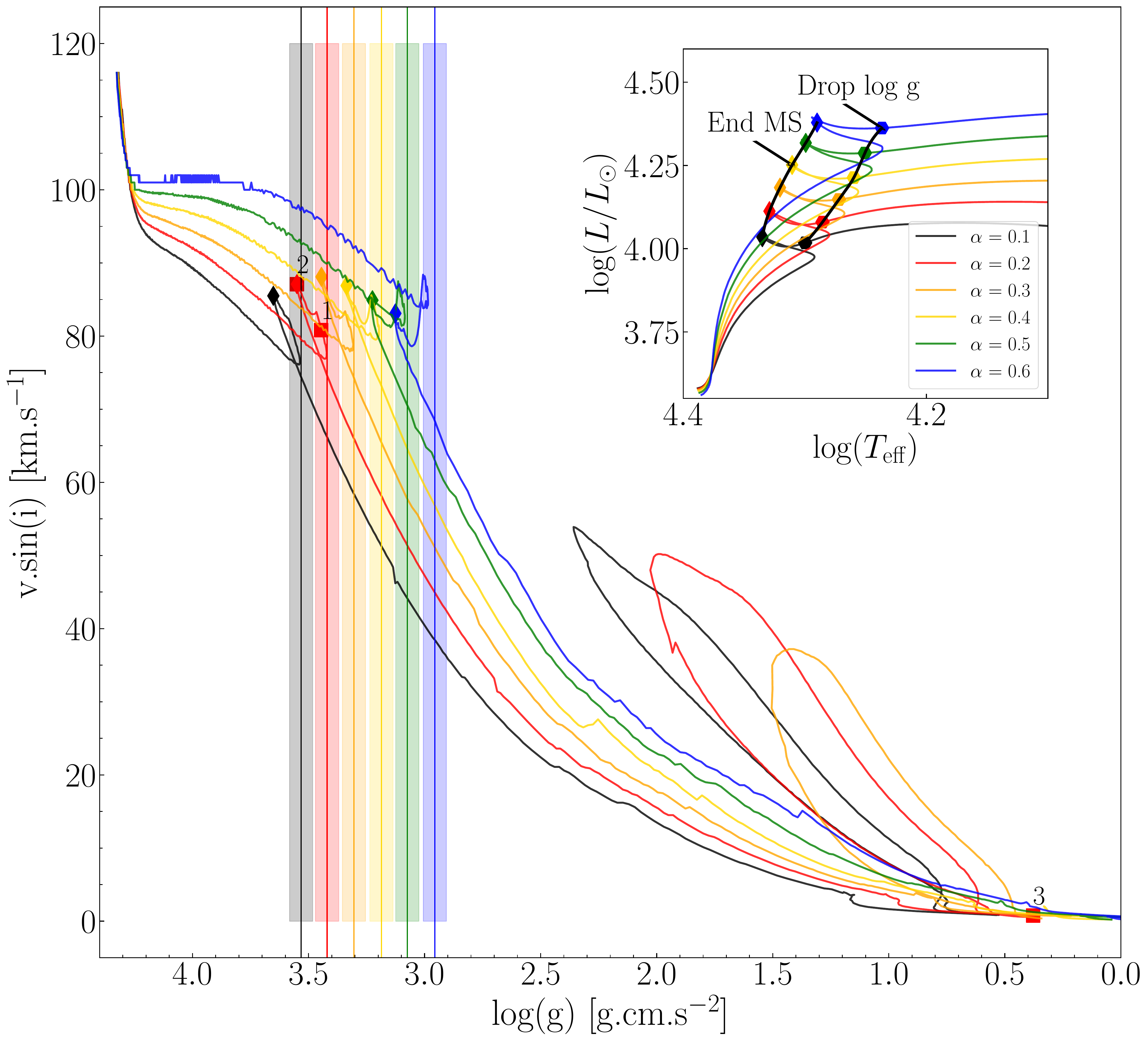}
    \caption{Evolution of the surface velocity as a function of the surface gravity for 9\Msol\ stars rotating at $\upsilon/\upsilon_\text{crit}=$0.2 with overshooting scaling from 0.1 to 0.6, computed with a moderate angular momentum transport. The vertical lines show the log(g) maximizing the ratio between the time spent on the left side of the shaded area and the time spent on the right side, computed with Eq.~\ref{Eq:gLim}. Diamonds show the end of the MS. Numbers associated with red squares are separated by 10$^5$ yrs from one another, with square (2) being the end of the MS. In the upper right corner is the HRD of the same models, with the location of the velocity drop indicated. The drop occurs nearly at the same position as the red hook and thus this feature can be used to observationally detect its position.}
    \label{fig:vsini_vs_log_g_limit}
\end{figure}

Figure \ref{fig:vsini_vs_log_g_limit} shows the evolution of the surface velocity as a function of the surface gravity for 9\Msol\ stellar models computed with various overshoots and moderate angular momentum transport. After the initial very rapid drop of the surface velocity \citep[see discussion of this feature in][]{Denissenkov1999,StellarGRIDEkstrom2012,Granada2014}, the surface rotation varies slowly until the end of the MS phase, which occurs for $\log(g)$ between 3.5 and 3.0 in Fig.~\ref{fig:vsini_vs_log_g_limit}. After the MS phase, we see that the surface velocity drops strongly as the gravity decreases.
This is linked to the expansion of the envelope after the end of H-burning. The timescale of the drop is really short compared to the main sequence. Indeed, numbers associated with red squares on Fig. \ref{fig:vsini_vs_log_g_limit} are separated by 10$^5$ yrs from one another, with square (2) being the end of the MS, and yet the two first squares are separated by 0.1 dex in log(g), while the next two are separated by more than 3.0 dex in log(g). As noted above, this expansion is rapid and the surface velocity is likely dominated here by the local conservation of the angular momentum.
Observationally, Fig.~\ref{fig:vsini_vs_log_g_limit} means that very few {post-MS} stars will be expected in the high-velocity range (near the values at the end of the MS phase), but more will be observed at a much lower gravity when the core-He burning begins. 

We want to have an objective way of determining the log(g)-limit and comparing it among the various models. Since this limit is expected to mark the end of the slow-evolution MS phase and the transition to the rapid HR crossing, we use a consideration on the time spent before and after this limit.
In Fig.~\ref{fig:vsini_vs_log_g_limit}, the vertical lines show for each model the $\log(g_\text{lim})$ where the highest ratio between the time spent in the shaded area before and after the vertical line is (the black shaded area, for instance, corresponds to the case of the black track, and the same for the other colors). 
The position of $\log{(g_\text{lim})}$ is given by:
\begin{equation}
\label{Eq:gLim}
\log(g_\text{lim})= \text{Max.} \frac{t[\log(g_\text{lim})]-t[\log(g_\text{lim})+0.05\text{dex}]}{t[\log(g_\text{lim})-0.05]-t[\log(g_\text{lim})]}
,\end{equation}
where $t[\log(g_\text{lim})]$ is the age of the star when the surface gravity is equal to $g_\text{lim}$.

Here, we used 0.05 dex as the width to search for the maximizing limit, but lower and higher values have been tested and lead to the same results. Moreover, to exclude the possibility of finding a maximum ratio between two negligible times (compared to MS lifetime), we are searching for limits that maximize both the ratio and the total accumulated time. These limits show that indeed the drop in surface velocity occurs {at the same position as the red hook and thus can be used to determine the position of this feature.}



When increasing the overshooting, the limit shifts to lower log(g). Typically, passing from an overshooting $\alpha=0.1$ to 0.3, shifts the value of $\log(g_\text{lim})$ by 0.2 dex towards lower values. This is larger than the error in $\log(g)$ determinations from spectroscopy (about 0.1 dex).
Mass also plays a role in the impact of overshooting on the log(g) limit as Fig \ref{fig:maximize_vsini_mean_7and9_15and25_bin035.pdf} shows. We shall discuss this point in Sect.~4.

\section{Comparison with observations}
\label{section:Comparaison}

\subsection{Width of the main sequence band}
\label{section: main sequence band's width}
\begin{figure*}
    \centering
    \includegraphics[width=0.99\textwidth]{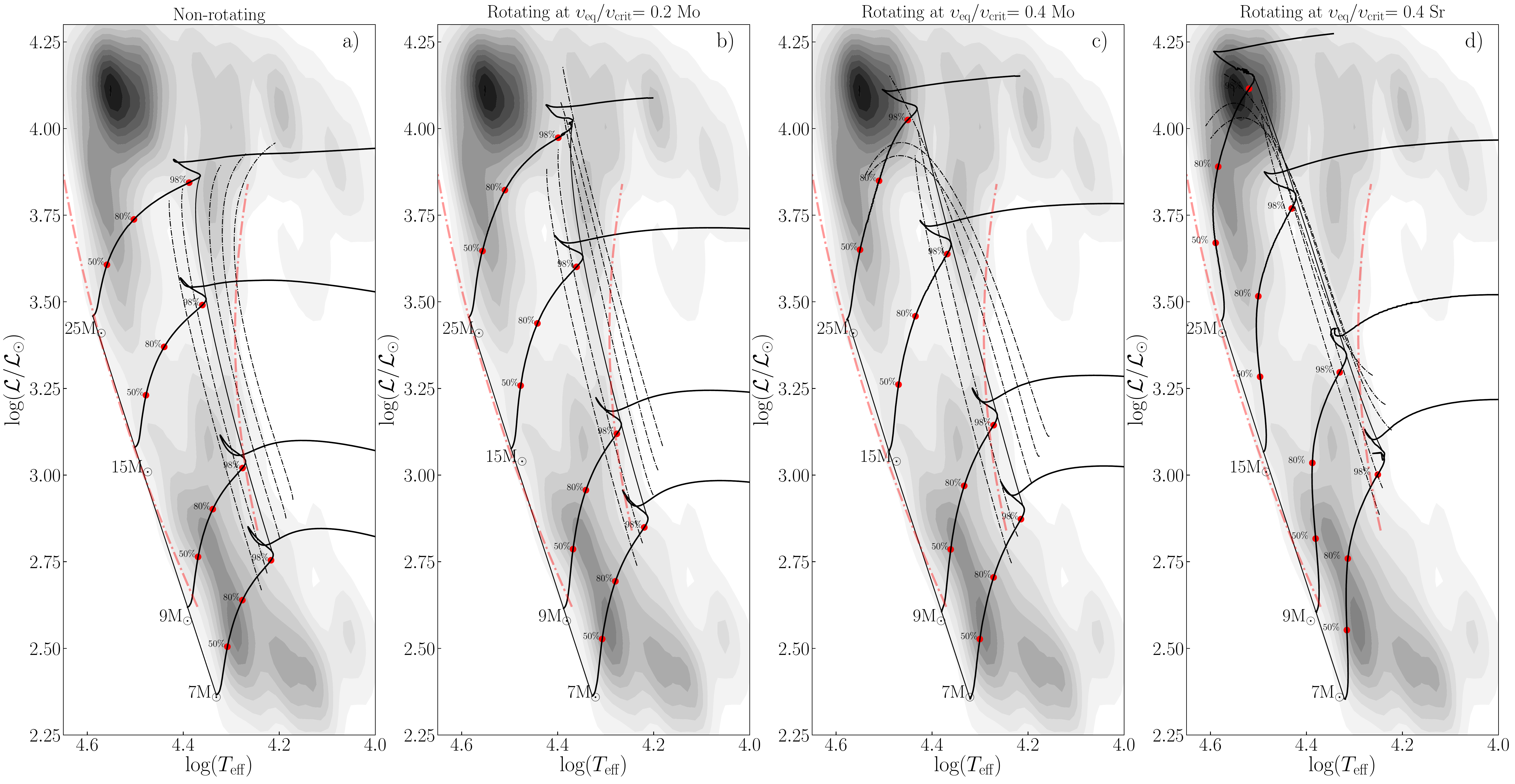}
    \caption{Spectroscopic Hertzsprung-Russell diagrams \citep[][$\mathcal{L}\equiv$ T$^4_\text{eff}$/g]{LangerKudritzki2014} for non-rotating, rotating at $\upsilon/\upsilon_\text{crit}$=0.2, and rotating at $\upsilon/\upsilon_\text{crit}$=0.4 with moderate and strong transport. The black {solid} lines connecting the red hooks of the tracks show the TAMS position for alpha=0.3, 
    the dashed lines show the TAMS line for overshoot parameters equal to 0.1, 0.2, 0.4, 0.5, and 0.6. The dashed-dotted red lines show respectively the empirical ZAMS (on the left in each panel) and the empirical TAMS limit (on the right in each panel) given by \citet{Castro2014}. The shaded areas show the kernel density estimation of stars from \citet{Castro2014} sample. It uses a Gaussian kernel, and the darker is the area, the higher is the proportion of the estimated underlying distribution that sits in that range.}
    \label{fig:HRD_power_plot}
\end{figure*}

Ideally, in order to establish the MS band's width, well-populated young clusters should be used since they provide the distribution of stars with a unique age and chemical composition. 
The fact that stars of the same age are observed makes that short phases of the evolution can be populated only by a very small range of initial masses. 
This is in contrast to the case of a mixed population of stars from different regions of the Galaxy (including, e.g., clusters of different ages and stars from the field), where the effect of different types of observational biases on the compiled sample blur the exact location of the end of the MS. 
In brief, the interpretation of such a survey is more complex than single-aged stellar populations in stellar clusters.

The stellar clusters present, however, their own difficulties. First, there are not so many very well-populated and well-observed stellar clusters at solar metallicity with turn-off masses in the range of interest here. 
Second, careful studies aimed at discriminating stars belonging to the clusters from the field stars need to be made \citep[here, \textit{Gaia} data will be extremely useful; see, e.g.,][]{Berlanas2020,deBurgos2020}. 

Another difficulty that pertains to any survey (field stars or stellar clusters) is the fact that photometry alone is not sufficient for the purpose of the present work.
Indeed, in many color-magnitude diagrams, the upper main sequence of young stellar populations is nearly vertical. This implies that tiny differences in colors may correspond to large changes in the effective temperature. Spectroscopic data are needed in addition to photometric data in order to make comparisons with theoretical models. 


In the present work, we use three compilations of empirical data for a large sample of Galactic O- and B-type stars located within a few kpc of the Sun. 
These three compilations are those by \citet{Castro2014}, \citet{SimonDiaz2017}, and \citet[][in prep.]{Holgado2020}.
We refer the reader to the above-mentioned papers for a detail description of the characteristics of the various  samples considered. 

The present theoretical tracks in Fig.~\ref{fig:HRD_power_plot} are plotted over the density distribution resulting from these empirical data in the spectroscopic HR diagram. Following a similar approach as in \citet[][]{Castro2014}, the shaded areas show the Gaussian kernel density estimation. The darker the area, the higher is the density of observed stars.


As explained in Sect.~3, a sharp drop in the number of observed stars is expected beyond the MS band.
\citet{Castro2014} have deduced the position of the ZAMS and of the TAMS (corresponding to the line joining the red hook discussed in Sec.~3.2) from their data. These lines are indicated as dashed-dotted red lines in each of the panels of Fig.~\ref{fig:HRD_power_plot}. 
Empirical ZAMS lines nearly perfectly match with the theoretical ones \citep[see, however, the discussion in][on the case of stars with masses above 30--40\Msol]{Holgado2020}.
The situation is clearly different for the TAMS lines. 
We can note the following points:

Whatever the rotation or angular-momentum transport efficiency considered, there is no one unique value of the overshooting parameter that can fit the empirical limit over the whole mass domain. The empirical limit does not follow the same slope as the lines connecting the red hooks of constant overshooting models. To fit the empirical line, it is necessary to consider values of $\alpha$ that increase with the initial mass, which is in agreement with the conclusion drawn by \citet{Castro2014}. 
For the non-rotating models, we get values of $\alpha$ increasing from 0.2 for the 7\Msol, to 0.3 for the 9\Msol, and 0.6 for the 15 and 25\Msol.
 
For the rotating models with $\upsilon/\upsilon_\text{crit}=0.2$ (moderate and strong transport), and the $\upsilon/\upsilon_\text{crit}=0.4$ with a moderate angular-momentum transport, we have $\alpha$=0.1, 0,2, ~0.6 for the 7, 9 and 15\Msol. The rotating 25\Msol\ has a high luminosity, which brings it above the empirical limit drawn by \citet{Castro2014}, hence, no $\alpha$ value can be given for this model.

For the models with $\upsilon/\upsilon_\text{crit}$=0.4 and a strong angular-momentum transport, we have $\alpha$=0.1 for the 7\Msol. For the 9\Msol\ and above, increasing the overshoot does not tend to produce larger MS bands -- in fact, it may even reduce it. Thus, for these models, the MS band reaches a maximum width for a given value of $\alpha$ and then decreases when $\alpha$ continues to increase. If the maximum width is bluer than the empirical limit, then there are no longer any possibilities for such models to fit this limit by increasing the overshoot.

The discussion above deals with individual comparisons of  the different models according to their rotation, but actually, when we are observing a population of stars, we have a mixture of initial rotations, and we may even have different efficiencies of angular momentum transport in case this efficiency depends on some initial conditions as the presence of a magnetic field, or favorable conditions for activating a dynamo. Thus in a stellar population, 
the non-rotating (or slowly rotating one) would follow track of the panel {\it a}, moderate and fast-rotating stars would follow the cases shown in the 
panels {\it b} and {\it c} (even higher initial rotation could have been considered but with these three initial rotations we already cover a great majority of the
observed surface velocities).

Keeping this point in mind, we consider what values of $\alpha$ would allow the best fit of the empirical limit. In the following discussion, we give a larger weight 
to the values of $\alpha$ resulting from panels {\it b} and {\it c} of Fig.~\ref{fig:HRD_power_plot} because these models well cover the range of observed surface rotation velocities. The models with $\upsilon/\upsilon_\text{crit}=0.2$ with a strong angular momentum transport gives similar results to those with a moderate angular momentum transport shown on panel {\it b}. Thus, at least for moderate rotations, whatever a moderate or a strong angular momentum coupling is considered, the same values of $\alpha$ would be deduced. 
For the rotating 7\Msol\ (whatever the panel we consider in Fig.~\ref{fig:HRD_power_plot}), we get that the best $\alpha$ value would be around 0.1.
This value is the same value as the one derived by \citet{StellarGRIDEkstrom2012} based on the observed MS band's width in a mass domain around 2\Msol\footnote{The physic inputs of the models by \citet{StellarGRIDEkstrom2012} is exactly the same as the one used in our models with
a moderate angular momentum transport.}. In the latter mass domain, stars are generally slow rotators, thus, the MS band width can reasonably be mainly attributed to the overshooting process alone. We find here that, up to a mass around 7\Msol, that is, well in a domain where it is more common to find stars with rotational velocities up to 400\,--\,450~\kms\ (i.e., $\upsilon/\upsilon_\text{crit}\geqslant$ 0.4), a value of $\alpha$ equal to 0.1 can be chosen (at least for the type of stellar models that we are investigating in the present work). 
We note that such a value would give too small a band width for the non-rotating models, but this is not a problem because the non-rotating models are not representative of all the stars.

Around the 9\Msol, $\alpha$ should be increased to a value of 0.2 (look at the two middle panels). In case of a very strong angular-momentum transport and for the models with
$\upsilon/\upsilon_\text{crit}=0.4$, much larger values of $\alpha$ should be used. However, since for more moderate rotation with a strong angular momentum transport, a value of 0.2 would fit, we shall stick to that value in that mass domain.

For the 15\Msol, the two middle panels would favor a value of $\alpha$ around 0.6. This would also be the case for the non-rotating (or very slowly rotating models). The fast-rotating ones with a strong overshoot will never populate the region near the empirical limits, thus, these models cannot be representative of the bulk observed populations. 

For the 25\Msol\ rotating models, the models at the end of the MS phase are in a range of luminosity beyond the region where the empirical limit is given. This is a consequence of the fact that rotation considerably increases the degree of mixing in these models and thus makes the tracks luminous and blue in the HR diagram. Thus, no overshooting value can be deduced for this model from the available observations. In a mass range around 25\Msol, only initially slowly rotating models with a large overshooting, around 0.6, can populate the region where the empirical limit of \citet{Castro2014} is. 

In conclusion, based on the considerations above, we deduce that the dependence of $\alpha$ with the initial mass would be 0.1 for the mass range between 2 and up to 7\Msol, around 0.2 for 9\Msol\ and 0.6 for 15 and 25 M$_\odot$.
In the following section, we investigate whether such a variation of $\alpha$ with the mass is compatible {}with the velocity drop feature.

\subsection{Surface velocities}
Before comparing the models with the observations, it is useful to first discuss synthetic populations \citep[computed with {\sc{Syclist}},][]{syclist}, as shown in Fig.~\ref{fig:Vsini_logg}. Although the results are model-dependent, they may give a few very useful guidelines for interpreting the observations. 

\subsubsection{Synthetic population}

\label{section:Synthetic Population}

\begin{figure*}
     \centering
      \includegraphics[width=0.48\textwidth]{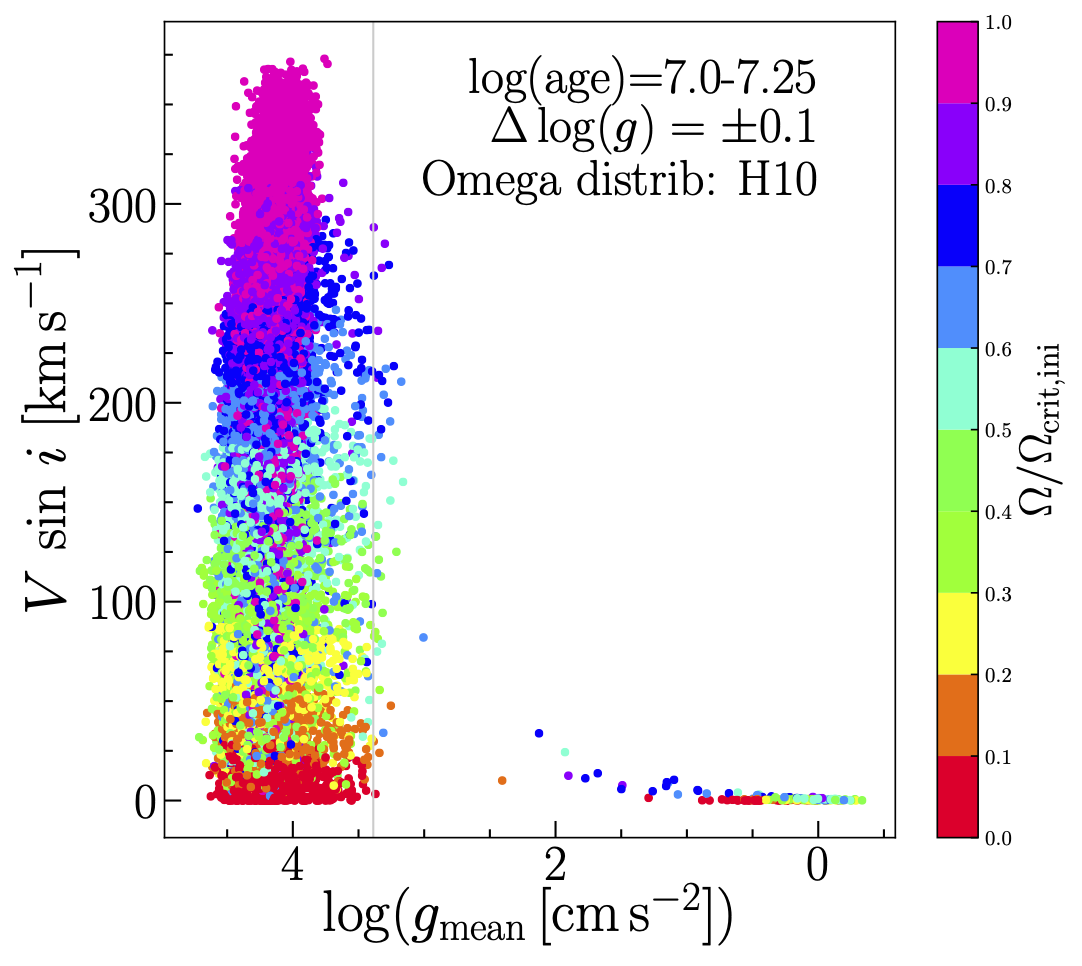}\includegraphics[width=0.48\textwidth]{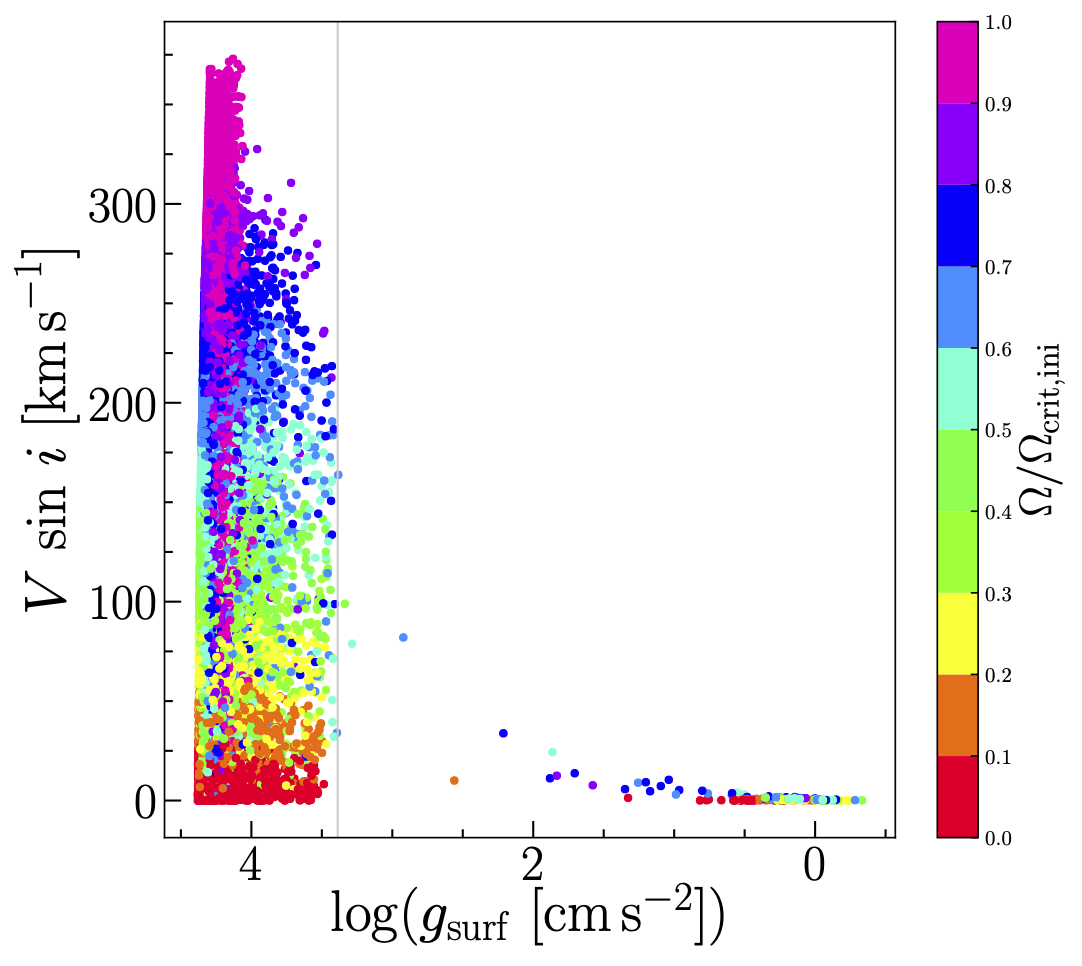}
     \caption{Synthetic populations of stars obtained by adding five clusters \citep[computed with {\sc Syclist},][]{syclist}, each of 10 000 stars at solar metallicity with a turn off mass around 15\Msol. The stellar models are from \citet{grids1p7_15_2013}. These models have the same physic inputs as the one used for computing our moderate angular momentum transport models with an $\alpha=0.1$. The range of ages spanned by the five clusters is indicated in the left panel. A noise of $\pm$ 0.1 dex is considered in $\log(g)$. An initial distribution of rotations as given by \citet{Huang2010} is used. The left panel shows a population with the $\log(g)$ limit determined for our 15\Msol\ model with $\alpha_\text{ov}=0.1$. The right panel shows the same population but this time the correction for the angle of view is not applied on $\log(g)$.} 
     \label{fig:Vsini_logg}
\end{figure*}

{Figure~\ref{fig:Vsini_logg} show the distribution of a synthetic population of 40000 stars in the $\upsilon\sin i$ versus $\log g$ diagram, with ages distributed between 10 and 18 My. 

To construct this diagram, stellar models with an $\alpha$=0.1 were used. The initial masses and rotations are randomly chosen so that the final distributions of the masses and rotations are equal to, respectively, the Salpeter's Initial Mass function and the distribution of the initial rotation velocities as deduced from observations by \citet{Huang2010}. We assumed a uniform distribution of the ages between the limits indicated above and a uniform distribution 
of the inclinations of the rotational axis with respect to the line of sight. 
Inclination effect changes the perceived color and flux received from a rotating star. A fast-rotating star appears brighter and bluer at the pole than at the equator. This has an impact also on surface gravity.
The plotted gravity is the flux weighted ($\sim T^4$) average of the local effective gravity on the visible hemisphere.
A star seen equator-on has a surface gravity lower than the whole surface flux averaged gravity, while a star observed pole-on will show a surface gravity that is stronger than the whole surface flux-averaged gravity.}

Such plots tell us how the transition from the MS band to the more evolved stage would appear if stars in a limited range of ages would be considered. 
Looking at the width in $\log(g)$ of the MS band (i.e., stars with a surface $\log(g)$ larger than around 3.4), we see that for $\upsilon\sin(i)$ values larger than 300 km s $^{-1}$ (the magenta dots), the MS band becomes narrower. This is a consequence of rotational mixing which, at such high rotation, is strong enough to keep the stars more compact and thus with a higher surface gravity during the MS phase.

After the MS band, there is a lack of stars in a domain between about 2.5 and 3.5, indicating that
the model stars evolve very rapidly in that gravity domain. This phase corresponds to the crossing of the HR gap. Then we can see again stars with low gravity ($\log(g) < 2.5$). These stars are burning helium in their core. The position in the HRD where the core He-burning begins depends on many ingredients of the models. In the present models, the core He-burning phase begins in the red supergiant phase. Considering lower metallicities or using different prescriptions for the rotational mixing may change this picture keeping stars with a relatively high surface velocity beyond the end of the MS phase.

{Considering these synthetic populations as mock observations, the value of $\log g$ at which the drop in velocity occurs might be taken equal to $\sim$3.4 if we
take the limit given by stars with $\upsilon\sin i$ below about 100 km s$^{-1}$, or equal to 3.1 if we take the limit given by stars with $\upsilon\sin i$ between 100 and 200 km s$^{-1}$. 

For the ages of the synthetic clusters considered here, the initial masses of the stars at the end of the MS phase are near 15 M$_\odot$. 
In the left panel of Fig.~\ref{fig:Vsini_logg}, we plot, as a vertical line, the limit of the MS band for a 15 M$_\odot$ computed with
an $\alpha$=0.1, an initial rotation equal to $\upsilon/\upsilon_\text{crit}$=0.4, and a moderate angular momentum transport. These properties correspond to the properties of the models used to construct the synthetic population and thus the vertical line indicates the true end of the MS band in this diagram.
Looking at the right panel of Fig.~\ref{fig:Vsini_logg}, which shows the distribution of stars for the same populations as the one shown on the left diagram, but with no inclination effect, we see that this limit perfectly matches the end of the MS band. 

The best strategy to obtain the position of the true drop off (here at $\log g$=3.4) is to take the limit given by the lowest gravity at which stars are
more or less uniformly distributed among low and high $\upsilon\sin i$ values.
Looking at the left panel of Fig.~\ref{fig:Vsini_logg}, we see that for $\log g$
above 3.4, such a uniform distribution of $\upsilon\sin i$ is realized, while below that value, a non-uniform distribution is obtained in the low velocity range ($\upsilon\sin i < 100$ km s$^{-1}$). 
For higher $\upsilon\sin i$, some points appear beyond this limit.}
This dispersion is mainly caused by two effects, first the impact of rotation on the tracks and second the inclination effect. For moderate rotation rates, tracks end their MS phase with a slightly lower surface gravity than non-rotating tracks. This is because in this velocity range, rotation produces more massive convective cores, while not driving an overly  strong mixing in the envelope that would keep the star more compact. 
The effect of inclination comes from the fact that a star seen equator-on would show a surface gravity that is smaller than the whole surface flux-averaged gravity.

\subsubsection{Comparison with IACOB project observations}

\label{section:surface velocity observations}
 \begin{figure}
     \centering
     \includegraphics[width=0.49\textwidth]{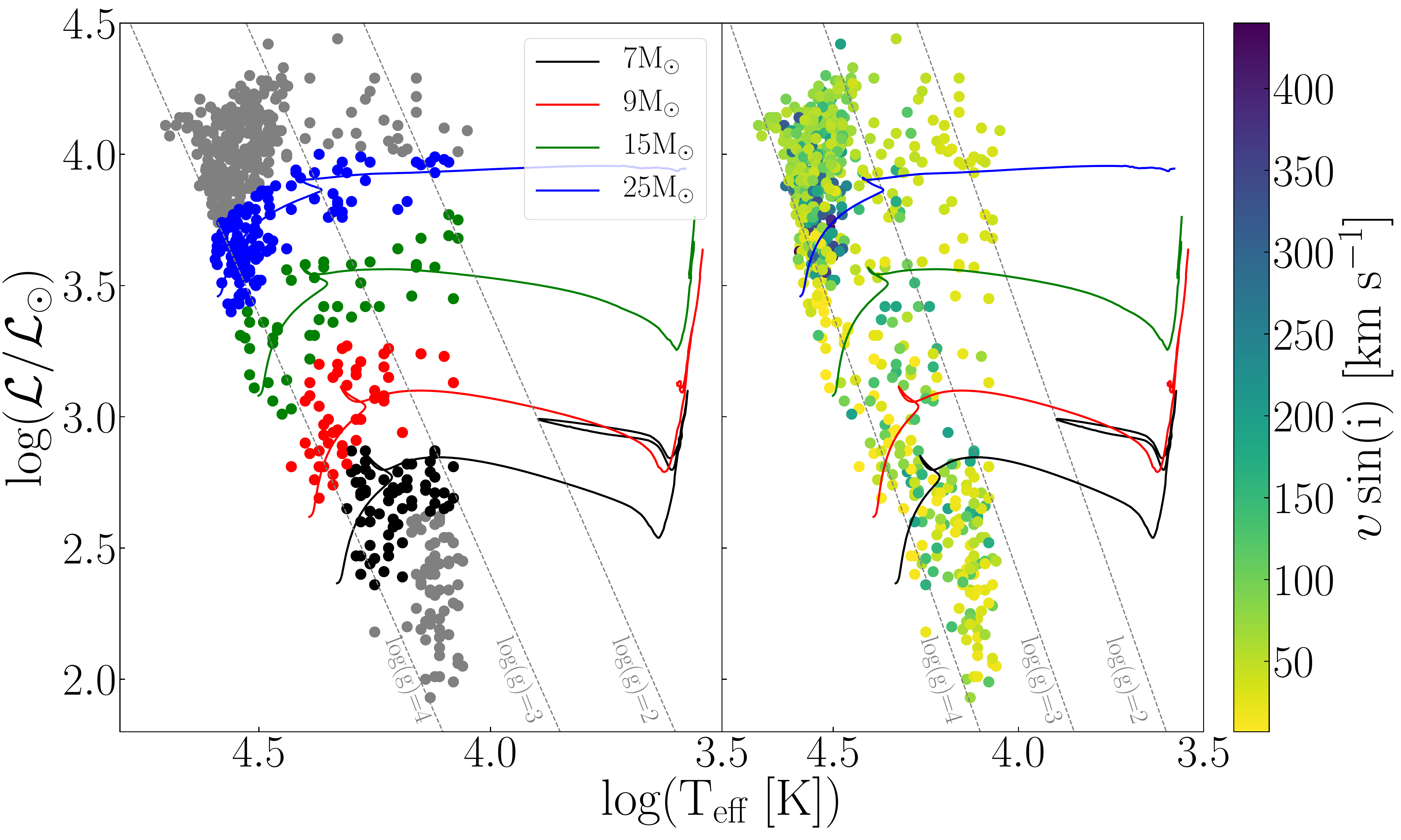}
     \caption{Positions of the stars with a determined $\upsilon\sin(i)$ taken from \citet{SimonDiaz2017} and \citet[][in prep.]{Holgado2020} in the spectroscopic HR diagram. The evolutionary tracks correspond to non-rotating tracks with an $\alpha$=0.3. The left-panel shows the mass selection and the right panel shows the $\upsilon\sin(i)$ distribution.}
     \label{fig:mass_selection}
     \end{figure}

  \begin{figure*}
     \centering
     \includegraphics[width=0.98\textwidth]{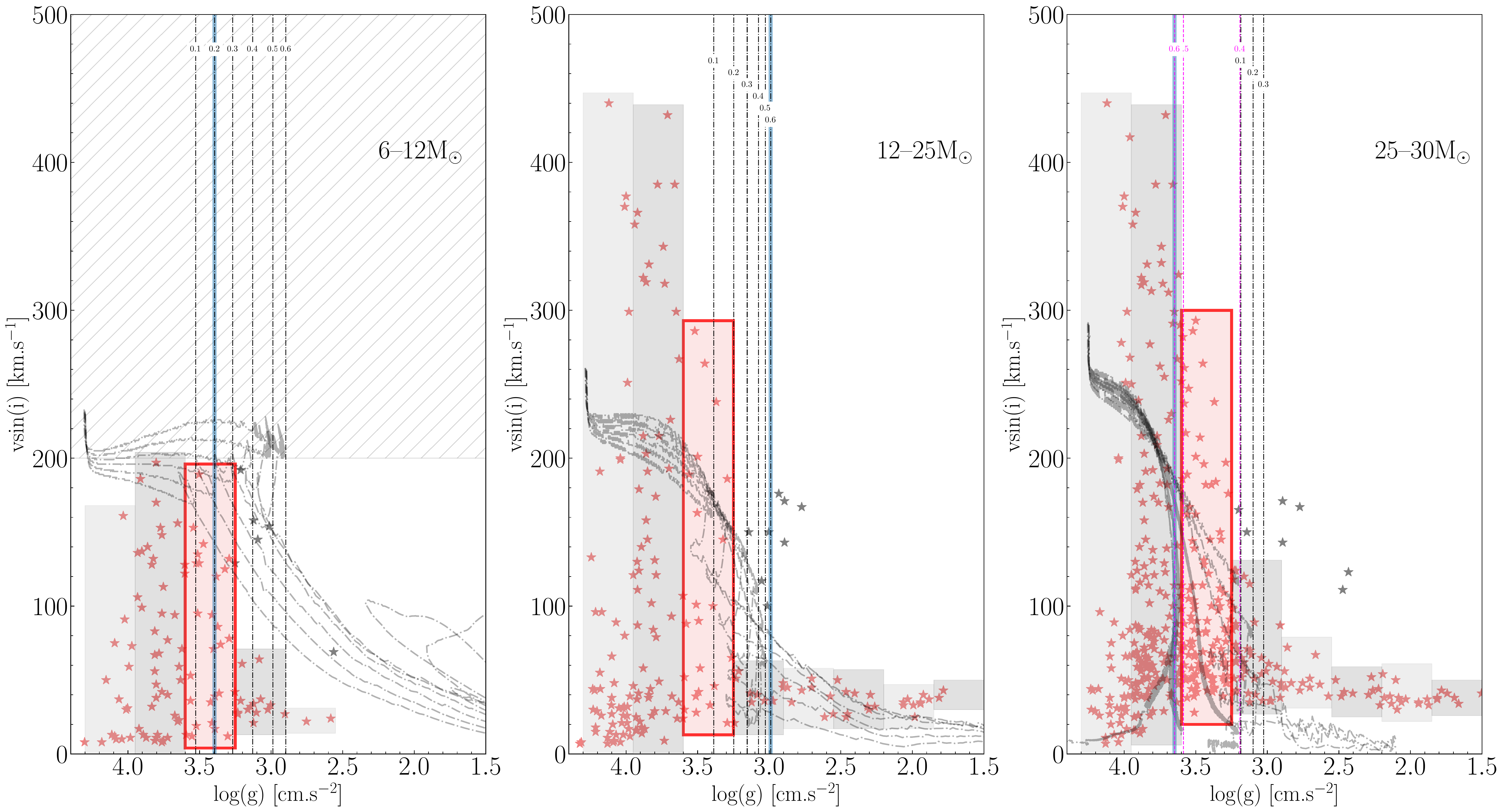}
     \caption {{Comparison of observed and predicted surface velocities as a function of surface gravity.} In each panel, the light grey dashed tracks show the evolution for a given initial mass model of the surface equatorial velocity as a function of the surface gravity, with overshooting equal to 0.1, 0.2, 0.3, 0.4, 0.5, and 0.6 (from left to right, except for the right panel, where models, due to blueward evolution at $\alpha \geq 0.4$, goes as 0.6,0.5,0.4,0.1,0.2,0.3 from left to right). The models have been computed with $\upsilon/\upsilon_\text{crit}=0.4$ and a moderate angular momentum transport. Models with an initial mass equal respectively to 9, 15, and 25\Msol\ are plotted in the left, middle, and right panel.
     The vertical lines shows for each overshooting value,  the value of the log(g) where the expected $\upsilon\sin(i)$ drop is predicted to occur. The one shadowed in blue shows the case corresponding to the model with the overshoot deduced from the MS band's width (see Sect. \ref{section: main sequence band's width}). 
     The colored stars are observed $\upsilon\sin(i)$ given by \citet{SimonDiaz2017} catalogue and brand new O-star sample with high $\upsilon\sin(i)$ from Holgado et al. (in prep.). For the mid/late B-stars, \citet[][]{SimonDiaz2017} excluded from their sample all stars with $\upsilon\sin(i) > $200\kms\ to study macroturbulent broadening (hatched zone in the left panel) . In each panel, we plot only those stars showing a position in the HR diagram indicating that they have a mass near the one of the plotted stellar models. 
     In each panel, the red band indicates the region where the $\upsilon\sin$(i) drop likely occurs. To determine its position, we use both observations and considerations based on the mock observations shown in Fig. \ref{fig:Vsini_logg} (see text). In that process, however, we do not take into account the grey stars that, according to Holgado et al. \citep[in prep.; see Fig. 1 in][]{Britavskiy2020}, might be the result of the evolution of an interacting binary system.}
     \label{fig:maximize_vsini_mean_7and9_15and25_bin035.pdf}
 \end{figure*}

The analysis of the sample used here, obtained within the framework of the IACOB project, is described in \citet{SimonDiaz2017}.
With the aim to provide new empirical clues about macroturbulent spectral line broadening in O- and B-type stars to evaluate its physical origin, Sim\'on-D\'iaz et al. (2017) compiled high-quality spectra of a sample of $\sim$430 stars with spectral types in the range O4\,--\,B9 (all luminosity classes). By means of a detailed quantitative spectroscopic analysis, these authors determined estimates of the effective temperatures (T$_\text{eff}$), surface gravity ($\log(g)$), and projected rotational velocity ($\upsilon\sin(i)$). For the sake of including relevant information for interpreting the diagrams presented in this section, we note that the sample of stars investigated by Sim\'on-D\'iaz et al. (2017) was limited to cases not identified as double-line spectroscopic binaries (SB2) and having a $\upsilon\sin(i)<$200 \kms.

The sample of Galactic OB stars considered by Sim\'on-D\'iaz et al. (2017) was recently updated with a sample of (285) likely single O-type stars (this time including the full range of $\upsilon\sin(i)$ values). In this case, spectroscopic parameters were presented in Holgado et al. (2020), and $\upsilon\sin(i)$ results are soon to be published (Holgado et al., in prep.)\footnote{Results for half the sample are already available in \citet{Holgado2018}.}.
Again, for a better understanding of the diagrams presented in this section, we remark that the sample of stars investigated by Holgado et al. (having all of the masses above 20\Msol) covers the full range of $\upsilon\sin(i)$ values (up to $\sim$450 \kms).

The complete sample of stars is plotted in a spectroscopic HR diagram in Fig.~\ref{fig:mass_selection}. The sample only populates the region from the ZAMS down to log(T$_\text{eff}) \sim$ 4.1 since this is the domain in the sHRD where O and B stars -- the main type of targets surveyed by the IACOB project -- are located.

Figure \ref{fig:maximize_vsini_mean_7and9_15and25_bin035.pdf} shows the surface velocity versus surface gravity diagrams for rotating models for three different bins in mass. In all cases, we also include model predictions for models with various $\alpha$ values for a characteristic mass, and those empirical results corresponding to subsamples of stars fulfilling the mass range criteria (see Fig. \ref{fig:mass_selection}).
The leftmost panel of Fig. \ref{fig:maximize_vsini_mean_7and9_15and25_bin035.pdf} contains all stars shown in black and red in Fig.~\ref{fig:mass_selection}, that is, those aggregating around the 7 and 9\Msol\ models. 
We overplotted the tracks for our 9\Msol\ with $\upsilon/\upsilon_\text{crit}$=0.4 (moderate angular momentum) and different overshoots. 
The surface velocity for the track is the actual (non-projected) equatorial velocity. For the observed points, $\upsilon\sin(i)$ is indicated.
This implies that the positions of the observed stars whose actual surface velocity would be equal to that of the model are on or below the tracks. 
The middle panel shows data points corresponding to higher initial masses (green and blue points in Fig.~\ref{fig:mass_selection}). The tracks are those of the present work for 15\Msol\ (otherwise the same physics inputs as for the tracks shown on the left panel). The right panel shows points corresponding to even higher masses (blue points and grey points above) overplotted on 25\Msol\ tracks (same physics inputs as other panels).

In the observed sample, we do not expect to see the sharp drop in number beyond the MS phase. Indeed looking at Fig.~\ref{fig:mass_selection}, we do not see any sharp drop in the number of stars in the effective temperature range covered. Thus, here the feature that is to be compared with the theoretical prediction is the drop in surface velocity, not the drop in the number of stars. In this regard, we remark again that, as indicated above, the sample of stars considered in the leftmost panel of Fig. \ref{fig:maximize_vsini_mean_7and9_15and25_bin035.pdf} only includes stars with $\upsilon\sin(i)$<200 \kms, an empirical bias which also partially affects the middle panel, but not the rightmost panel. 

For the purposes of comparison, we binned the data points.
The width of the bins in logg (here 0.35 dex) is a compromise between reaching a sufficiently good resolution in $\log(g)$ and at the same time having a sufficient number of points in each bin to appropriately derive a meaningful mean value for
$\upsilon \sin i$.
We tested other widths of the bins and obtained the same results as those quoted here. 

We colored in blue the vertical line showing the end of the MS band corresponding to the value of $\alpha$ as deduced from the discussion in Sect. \ref{section: main sequence band's width}. In the following, we ignore the grey stars for determining the $\upsilon\sin(i)$ drop since according to de Burgos et al. (in prep.), these stars might result from the evolution of interacting binary systems. 

As discussed in Sect. \ref{section:Synthetic Population}, 
 as the end of the MS band we chose the last bin before the observed $\upsilon\sin(i)$ drop, where the $\upsilon\sin(i)$ are still distributed continuously over a large range. In each panel of Fig. \ref{fig:maximize_vsini_mean_7and9_15and25_bin035.pdf}, we highlighted  this bin in red. 

{First,} in the case of stars with masses around 6--12\Msol\ (left panel of Fig.~\ref{fig:maximize_vsini_mean_7and9_15and25_bin035.pdf}){,} the third bin (from the left) seems like the most appropriate choice with this method. The value derived from the MS width band is compatible with this result. 
This leads to the conclusion that in the  6--12\Msol\ mass range, an $\alpha$ between 0.1 and 0.3 seems appropriate.

{For} the case of the stars with higher initial masses, between 12--25\Msol\ (see the middle panel of Fig.~\ref{fig:maximize_vsini_mean_7and9_15and25_bin035.pdf}){,}
using the same method, we obtain an overshooting $\alpha$ between 0.1 and 0.2 (see the bin highlighted in red). These values are thus smaller than those derived from the MS width.

Finally, let us consider the last mass range, 25--30\Msol\ (see the right panel of Fig.~\ref{fig:maximize_vsini_mean_7and9_15and25_bin035.pdf}). The 25\Msol\ models with $\alpha > 0.4$ enter into the Wolf-Rayet phase during the MS phase and thus show an evolution towards a higher surface gravity during the MS phase (see the first two tracks starting from the 
left in the third panel). Thus, a high $\alpha$ value already produces WR stars  over the duration of the MS phase from 25\Msol\ models with an initial rotation that is moderate. {This would likely  produce too many WR stars (from single stars) with respect to what has been observed. Thus, such high values of $\alpha$ do not appear reasonable and we do not consider them here.}
The third bin (from the left) seems an appropriate choice for the $\upsilon\sin(i)$ drop. Taken at face value, this would point either to an $\alpha$ value between 0.4 and 0.5 or to an $\alpha$ value around 0.1. If we discard the value of 0.5, then remain two possibilities, either 0.1 or 0.4.
At the moment, a conservative approach would be to stick to the value obtained in the 12-25\Msol\ range.

In summary, {for the mass range below $\sim$12\Msol, the surface velocity drop method gives a range of values for $\alpha$ compatible with the one given by the method based on the width of the MS band. For the mass range above $\sim$12\Msol,}
the values of the overshooting parameter $\alpha$ obtained from the $\upsilon\sin(i)$ drop are {lower than} the ones obtained from the MS width. 
{In this mass range, we currently favor the results given by the velocity drop over the one coming from the spectroscopic HR diagram.}
The main reason for this is that in the upper HR diagram (for masses above $\sim$ 12\Msol), the observations 
do not yet provide a well-defined position for the
drop in the density of stars.
 In light of all the above, we suggest a value of $\alpha$=0.1 for 7\Msol\ and $\alpha$=0.2 for masses between 9 to 25\Msol. {Interestingly, this choice appears to be confirmed by the results obtained by \citet{Tkachenko2020} using eclipsing binaries (see the discussion in Sect.~5).}

\subsection{Synthesis of these comparisons with the observations}

In summary, we obtain from the comparisons made {in the above sections} the following values of $\alpha$:

{
(1) $\alpha$ = 0.1 for masses between about 2 and 7\Msol;

(2) $\alpha$ = 0.2 for 9\Msol;

(3) For 15\Msol~and 25\Msol\ models, we suggest $\alpha$ = 0.2 as a conservative choice, but other options are possible as seen in the precedent sections. 

}

The values of $\alpha$ given above are only valid for the physic inputs of the considered models. The convective core sizes appear to increase more rapidly with the mass than given by models with a constant step overshoot $\alpha$=0.1. We find here a moderate increase of $\alpha$ with the mass occurring between 7 and 9 M$_\odot$.
In the next two sections, we first discuss the robustness and the limitations of these results and 
indicate some perspectives for future works in this  area of research.

\section{Discussion}
\label{section:Discussion}

{
\subsection{Limitations of the different methods}

Each of the methods has its own specific limitations. {At least two limitations are generic to all methods:  the first one,} is that whatever the observed feature considered, these features depend on more physical processes than just convection. Mass loss by stellar winds and transport in the radiative zones are, for instance, two important processes that also affect the position of the TAMS in the HR diagram and of the surface gravity at which the velocity drop 
occurs. The second one comes from the reliability of the parameters derived from the observed sample. 

\subsubsection{Width of the MS band}

For the MS width, one of the main limitations 
comes from the difficulty in the upper part of Fig.~\ref{fig:HRD_power_plot} to find the position of the end of the MS band as a clear drop of the stellar density. The situation is much better in the lower mass range. Thus the results obtained for masses between 7 and $\sim$15 M$_\odot$ are likely more reliable than those for higher masses. Another difficulty comes from the fact that unresolved binaries can actually populate an area beyond the TAMS predicted by single-star models \citep{Wang2020}. 
We note, however, that some of the stars in \citet{Castro2014} not identified as SB2 could still be SB1 systems or mergers. Hence, it is difficult to be completely sure that we have eliminated all binaries in the sample. The same applies to the two other samples \citep[][]{SimonDiaz2017,Holgado2020}.                                   
                                 
\subsubsection{Velocity drop}

In the observed sample used here, there is an important observational bias for the mid and late B-stars since \citet[][]{SimonDiaz2017} excluded from their sample all stars with $\upsilon\sin(i) > $200\kms\ to study macroturbulent broadening. Moreover, the B-MS stars are highly biased towards low $\upsilon\sin(i)$ stars due to their interest in determining abundances in the Solar neighborhood. In this case, we note that the sample for the 7-9\Msol\ would need to be more populated (both in high $\upsilon\sin(i)$ for MS stars, and low $\upsilon\sin(i)$ for post-MS stars) to better constrain the $\alpha$ value. The situation is better for the mass range of 15 and 25\Msol. 

Another major difficulty comes from the effect of binaries and of inclination that blurs the picture.  Stars resulting for instance from a merger may have a surface rotation very different from the one predicted by single star evolution.
{Inclination effect may populate regions beyond the end of the MS band with
fast rotating MS stars. In that case, the positions of these fast-rotating stars in the $\upsilon\sin i$ versus $\log g$ diagram do not necessarily require a larger convective core to be fitted but at least an appropriate account for the effects of inclination on their surface properties.}

In the present models, we used the same mass loss prescriptions as in \citet{StellarGRIDEkstrom2012}. These mass loss recipes account for the bistability effects. This effect, found by \citet{Vink2010}, results in a rapid and strong increase of the mass loss by stellar winds when some limits to effective temperatures are reached. If this happens before the end of the MS phase, it may produce a drop in the surface rotation that can be wrongly
interpreted as having been due to the expansion of the star at the end of the MS phase. 
We checked that the drop in velocity shown by our stellar models is due to the
expansion of the star after the MS phase and not to this bistability effect.  Such an effect can occur before the end of the MS phase in faster-rotating models than those
considered here. Typically stars should rotate faster than 70\% of the critical velocity.  Such fast rotators are very rare and thus cannot explain the drop seen in Fig.~\ref{fig:maximize_vsini_mean_7and9_15and25_bin035.pdf}.
}

\subsection{Impact of different physical ingredients for the models}

In the present work, we considered extending the convective core above the limit given by the Schwarzschild limit. {We may  wonder whether the results would be different if we had chosen to extend the convective core above the limit given by the Ledoux limit. During the MS phase, we do not expect any difference}. Indeed, since the core is decreasing in mass, there is no chemical composition gradient at the border of the convective core and, hence, there is no difference between the Schwarzschild and Ledoux criteria.
Differences appear, of course, during the core He-burning phase, where the convective core is increasing in mass. Thus changing from the Ledoux to the Schwarzschild criterion has an impact on the extension of the blue loops, on the changes of the surface abundances, and on the sizes of the intermediate convective zones \citep[occurring during the core He-burning phase, see e.g., the discussion in][]{georgy2014,Kaiser2020} but, of course, not on the observed features linked to the core H-burning phase. 

We consider here the extension of only the convective cores. {We could have considered applying an overshoot to all convective zones (intermediate, outer convective zones)}.
During the MS phase, there are no significant intermediate convective zones, nor convective envelope, thus our choice to discard models with overshooting for intermediate convective zones or convective envelopes is not a strong limitation of the present work. Intermediate convective zones and an outer convective zone, in general, appear during the core He-burning phase. 

In this work, we focused on step overshooting while other studies use a diffusive approach for the mixing of the chemical elements in the overshooting region. Depending on which approach is used, the gradients of chemical composition at the border of the core undergo modification: very steep gradients are expected for the step overshooting (if acting alone) and much smoother gradients in case of the diffusive approach. Asteroseismology can in some cases probe the structure of the chemical gradients at the border of the convective core and thus provide some hints favoring one over the other among these two approaches. We did not consider the diffusive overshooting type here but we do discuss rotating models that actually produce both an extension of the convective core (due to another physical process than overshooting) and a smoothing of the chemical gradients just above the core. Thus, to some extent, while it may be for a different physical reason, these rotating models cover (at least in a qualitative sense) the cases of models that would have been computed with a diffusive overshoot. We note however that since the physical process at work are different, the value for $\alpha$ that that would be deduced from non-rotating diffusive overshooting models and the values deduced from the present rotating step-overshoot models are likely to be different. 
On the other hand, diffusive overshooting models would likely also require an increase of the overshoot with the mass. This is indeed what we see when we look at our rotating models that require such an increase. 

In addition to the two cases of moderate and strong angular momentum transports considered here, many other different ways of accounting for the effects of rotation exist. A limited study of the impact of different prescriptions for the shear diffusion coefficient and the horizontal diffusion coefficient can be found in \cite{GM2013}. 
However, the cases discussed here ($\upsilon/\upsilon_{\rm crit}=[0,0.2,0.4]$, $\alpha$=[0.1,...,0.6] and Sr/Mo transport) explore more extreme situations in term of angular and chemical element transports than those explored in the above reference. {Thus, it is quite possible that present models cover all the situations
(and even more) resulting from these different prescriptions.}

So, on the whole, the values quoted at the end of Sect.~4 for $\alpha$ do not appear to be too much dependent on the fact that we chose the Schwarzschild instead of the Ledoux criterion, whether overshoot is implemented or not for modifying the extension of intermediate or external convective zone. They also do not appear to be too strongly dependent on the two types of angular momentum transport considered in this work (moderate or strong). This last point has to be taken with some caution. We have seen that the two types of models are similar for moderate initial rotation but may differ significantly for fast-rotating models.



{
\subsection{Comparisons with other models}

\begin{figure}
    \centering
\includegraphics[width=0.48\textwidth]{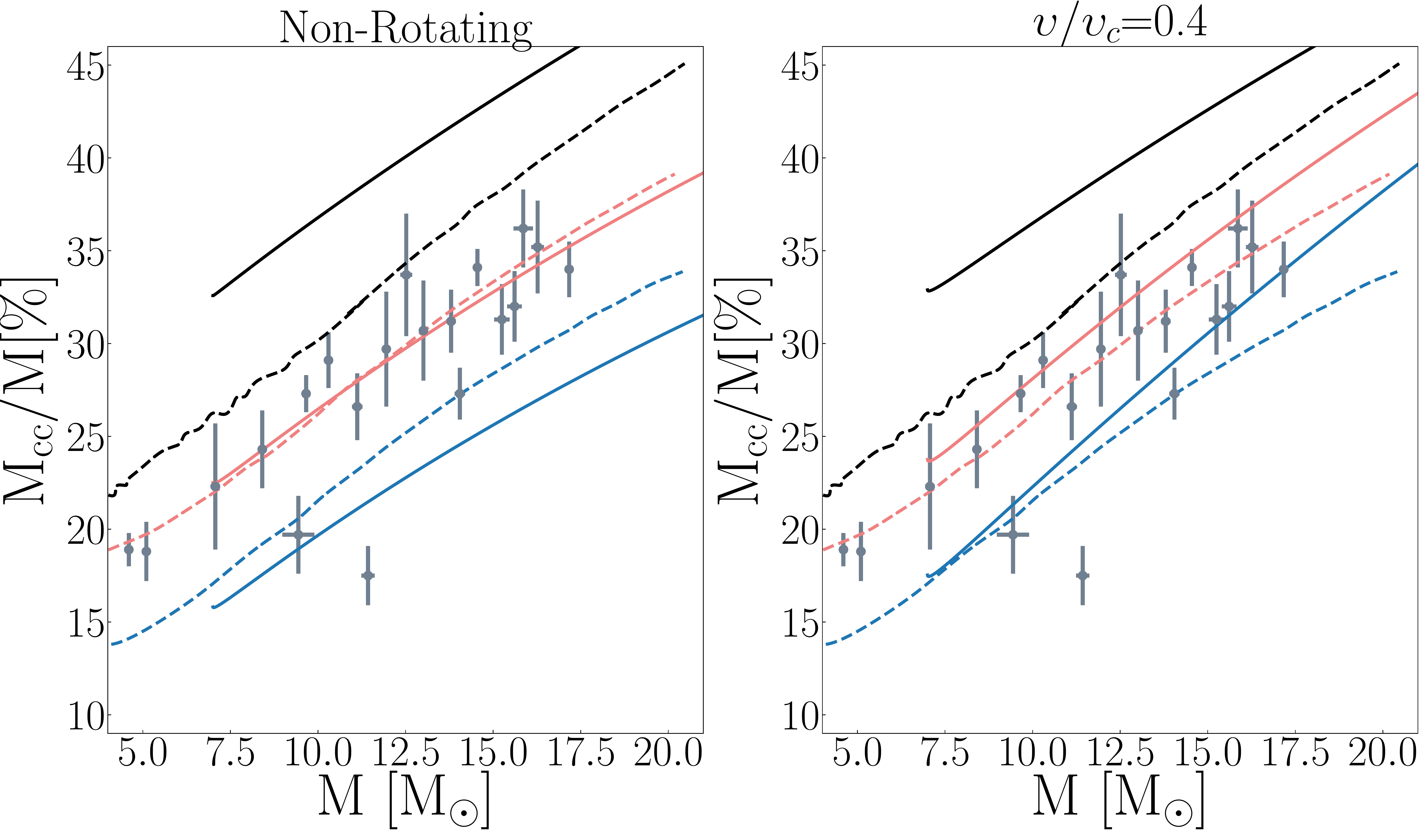}
\caption{{Variation of the convective core size at different stages during the core H-burning phase, as a function of the initial mass. {The dashed lines are the models of \citet{Tkachenko2020} with strong convective boundary mixing, the solid ones are the present GENEC models with $\alpha$=0.2. The tracks in black correspond to the ZAMS stage, while the red tracks correspond to a stage where the mass fraction of hydrogen in the convective core ($X_{\rm c}$) is 0.35, and the blue tracks to $X_{\rm c}$=0.1.}
The grey points show observed eclipsing binaries and their core mass estimates from \citet{Tkachenko2020}. 
In the left panel, the GENEC models are the non-rotating ones. In the right panel, they are the $\upsilon/\upsilon_{\rm crit}$=0.4 rotating ones.}}
\label{Tka}
\end{figure}

We make use of the models discussed in \citet{Tkachenko2020} to briefly make a comparison between the core sizes they obtain with those presented here. These authors have used 11 eclipsing binaries with masses of the components between 4 and 16 M$_\odot$ to determine the convective core size during the core H-burning phase. 
In Fig.~\ref{Tka}, we compare the size of the convective cores obtained at the middle of the MS phase with those obtained by \citet{Tkachenko2020}.
We see that the line corresponding to our non-rotating model with $\alpha$=0.2 nearly perfectly matches their line corresponding to a strong convective boundary mixing, which is their favored model. For the rotating models, models with $\alpha$ between 0.1 and 0.2 would match their line, which is obtained with a strong boundary mixing.  Interestingly, the slope of the variation of the core mass with the mass is very similar in models with a constant step overshoot and the ones with a convective boundary mixing as implemented in \citet{Tkachenko2020}. These comparisons illustrate the fact that comparing predictions of stellar models with observations as those above cannot constrain the mixing at the boundary of the core; however, they can provide insights on the size of the convective core, whatever the physical process responsible for it (boundary mixing, overshoot, rotation). We see here that adopting a value of $\alpha$ around 0.2 for masses
up to 16 M$_\odot$ would provide a good fit to the models, which,
according to \citet{Tkachenko2020}, best match the constraints from their sample  of eclipsing binaries.
}

\section{Conclusions}

The main findings of the present works are the following:
\begin{itemize}
\item {We discuss the impact of different overshooting values on two observed features and within the framework of three different types of models: non-rotating, rotating with a moderate angular momentum transport, rotating with a strong angular momentum transport.}
\item {A constant $\alpha$ value (here $\alpha$=0.1) for a step-overshoot between $\sim$2 and $\sim$ 7\Msol\ allows us to adequately fit the observed TAMS and the observed drop of the surface velocity.}
\item We confirm, as found by earlier works and other works in progress \citep[][]{Castro2014, Scott2021} that for more massive stars, there is a need to consider an $\alpha$ value that increases with the mass.
\item {The increase of $\alpha$ with the mass that we can deduce from comparisons of the present stellar models with the position of the TAMS and from the position of the velocity drop is not of the same amplitude. A larger increase is obtained from the MS band width. However, in the upper mass range, the TAMS position coming from the observations is not easy to determine due to the absence of a very clear drop in the stellar density. Thus, at the moment, we favor the lower increase given by the velocity drop feature. Such a choice appears to be supported when comparisons of our models are carried out with those of \citet{Tkachenko2020}.}
\end{itemize}

{The two features that we discuss in the present work are not the only ones that depend on the size of the convective core. We can include at least two additional features: the surface abundances reached at the end of the MS phase and the extension of the blue loops. Although we did not discuss these two features here, we checked that  $\alpha$ between 0.1 and 0.3 can all fit very well the N/H excess observed at the surface of Galactic MS B-type stars with initial masses inferior to 20\Msol\ \citep[see][]{Gies1992,Kilian1992,Morel2008,Hunter2009}. The models presented in this work feature blue loops crossing the instability strip for all the values of $\alpha$ below about 0.3 (see Fig. \ref{fig:Grid HRD}). Thus, in that respect, the values obtained from the discussions that precede (0.1 for the 7\Msol\ and 0.2 for the 9\Msol) are compatible with the presence of blue loops and of Cepheids. 
}

Among the more promising paths to progress along this line of research, we certainly see  improvements that are to be made in our understanding of the physics of turbulence both due to convection and induced by rotation. This is, however, a rather long-term project and one in which actual observations will still be needed to some extent in order to calibrate some parameters describing processes occurring in timescales and space-scales smaller than present-day resolutions of 3D simulations. 

A single-age population would bring stronger constraints on the problems discussed above, but the rarity of such clusters in the mass range studied here is a matter of concern. This emphasizes the importance of observational surveys such as the one performed by the IACOB project \citep{SimonDiaz2015} to converge  to attain a better understanding of massive star evolution.
\begin{acknowledgements}
The work by SM is supported by the project 200020-172505 interacting stars funded by the SNSF. GM, SE, CG, PE, SS have received funding from the European Research Council (ERC) under the European Union's Horizon 2020 research and innovation program (grant agreement No 833925, project $\text{\sc{Starex}}$). NC gratefully acknowledges funding from the Deutsche Forschungsgemeinschaft (DFG) - CA 2551/1-1. SS-D acknowledges support from the Spanish Government Ministerio
      de Ciencia e Innovaci\'on through grants PGC-2018-091\,3741-B-C22
      and CEX2019-000920-S, and from the
      Canarian Agency for Research, Innovation and Information Society
      (ACIISI), of the Canary Islands Government, and the European
      Regional Development Fund (ERDF), under grant with reference
      ProID2020010016.
\end{acknowledgements}


\label{section:Conclusion}

\bibliographystyle{aa}
\bibliography{aa}

\end{document}